\documentclass{article}
\def\1ad{\mbox{\normalsize $^1$}}
\def\2ad{\mbox{\normalsize $^2$}}
\def\3ad{\mbox{\normalsize $^3$}}
\def\4ad{\mbox{\normalsize $^4$}}
\def\5ad{\mbox{\normalsize $^5$}}
\def\6ad{\mbox{\normalsize $^6$}}
\def\7ad{\mbox{\normalsize $^7$}}
\def\8ad{\mbox{\normalsize $^8$}}
\def\makefront{
    \vspace*{1cm}\begin{center}
    \def\sp{
	\renewcommand{\thefootnote}{\fnsymbol{footnote}}
	\footnote[1]{corresponding author~~E-mail: \email_speaker}
	\renewcommand{\thefootnote}{\arabic{footnote}}
    }
    \def\newtitleline{\\ \vskip 5pt}
    {\Large\bf\titleline}\\
    \vskip 1truecm
    {\large\bf\authors}\\
    \vskip 5truemm
    \addresses
    \end{center}
    \vskip 1truecm
    {\bf Abstract:}
    \abstracttext
    \vskip 1truecm
}

\usepackage{times}
\usepackage{amssymb,amsmath,bbm}
\usepackage[dvips]{graphicx}
\def\a{\alpha}
\def\g{\gamma}
\def\eps{\epsilon}
\def\z{\zeta}
\def\h{\eta}
\def\th{\theta}
\def\m{\mu}
\def\r{\rho}
\def\p{\phi}
\def\vp{\varphi}
\def\j{\psi}
\def\Ga{\Gamma}

\def\sfrac#1#2{{\textstyle\frac{#1}{#2}}}
\def\rd#1{\buildrel{_{_{\hskip 0.01in}\rightarrow}}\over{#1}}
\def\ld#1{\buildrel{_{_{\hskip 0.01in}\leftarrow}}\over{#1}}
\def\+{\dagger}
\def\={\ =\ }
\def\pa{\partial}
\def\non{\nonumber}

\def\ch{\mathrm{ch}}

\def\tah{\mathrm{th}}
\def\e{\,\mathrm{e}\,}
\def\es{\,\mathrm{e}_{{}_{\star}}\!\!}
\def\im{\,\mathrm{i}\,}
\def\diff{\mathrm{d}}
\def\vel{\mathrm{v}}
\def\res{\mathop{\mathrm{res}}}

\newcommand{\C}{\mathbb C}
\newcommand{\R}{\mathbb R}
\newcommand{\unity}{\mathbbm{1}}

\newcommand{\Ecal}{{\cal E}}
\newcommand{\Lb}{\bar{L}}
\newcommand{\sta}{\star}
\newcommand{\se}{\sigma_1}
\newcommand{\sz}{\sigma_2}
\newcommand{\sd}{\sigma_3}

\newcommand{\cb}{}
\def\be{\begin{equation}}
\def\ee{\end{equation}}
\def\bea{\begin{eqnarray}}
\def\eea{\end{eqnarray}}
\begin {document}

\begin{flushright}      
hep-th/0409108\\        
ITP--UH--22/04\\        
\end{flushright}        

\def\titleline{
Noncommutative Sine-Gordon Model \footnote{
Talk presented at the XIIIth International Colloquium 
Integrable Systems and Quantum Groups in Prague 17-19 June 2004
and at the 37th International Symposium Ahrenshoop on the 
Theory of Elementary Particles in Berlin-Schm\"ockwitz 23-27 August 2004.}
}
\def\email_speaker{
{\tt 
lechtenf@itp.uni-hannover.de
}}
\def\authors{
Olaf Lechtenfeld
}
\def\addresses{
Institut f\"ur Theoretische Physik, Universit\"at Hannover\\
Appelstra\ss{}e 2, D--30167 Hannover, Germany\\
Email: lechtenf@itp.uni-hannover.de
}
\def\abstracttext{
As I briefly review, the sine-Gordon model may be obtained by dimensional
and algebraic reduction from 2+2 dimensional self-dual U(2) Yang-Mills
through a 2+1 dimensional integrable U(2) sigma model. I argue that the
noncommutative (Moyal) deformation of this procedure should relax the
algebraic reduction from U(2) $\to$ U(1) to U(2) $\to$ U(1)$\times$U(1).
The result are novel noncommutative sine-Gordon equations for a {\em pair\/}
of scalar fields. The dressing method is outlined for constructing its
multi-soliton solutions. Finally, I look at tree-level amplitudes to
demonstrate that this model possesses a factorizable and causal S-matrix
in spite of its time-space noncommutativity.
}
\large
\makefront


\section{Classical sine-Gordon model $\dots$}
Extremizing the sine-Gordon action 
\be
\cb S \= \sfrac12 \int\!\!\diff t\,\diff y \bigl[
(\pa_t\p)^2 - (\pa_y\p)^2 + 8\a^2 (\cos\p-1) \bigr]
\ee 
for a scalar field~$\p(t,y)$ on $\R^{1,1}$ with mass $=2\a$ yields
the sine-Gordon equation,    
\be \label{sG}
\cb (\pa_t^2 -\pa_y^2)\,\p \ +\ 4\a^2\,\sin\p \= 0\ .
\ee
This famous equation has many remarkable features, such as
a Lax-pair or zero-curvature representation, 
infinitely many conserved local charges,
a factorizable S-matrix without particle production,
as well as soliton and breather solutions.
The simplest soliton configuration (with velocity~$\vel$) is kink-like,
\be \label{kink}
\cb \p_{\text{kink}}(t,y) \= 4 \arctan\,\e^{-2\a\h}
\qquad\text{with}\quad \cb \h=\sfrac{y-\vel t}{\sqrt{1-\vel^2}} \ .
\ee
For later use I introduce light-cone coordinates
\be \cb
u\ :=\ \sfrac12(t+y) \ ,\quad v\ :=\ \sfrac12(t-y)
\qquad\Longrightarrow\qquad
\pa_u \= \pa_t + \pa_y \ ,\quad \pa_v \= \pa_t - \pa_y \ .
\ee

\section{$\dots$ via dimensional and algebraic reduction}
In 4d Yang-Mills and 3d Yang-Mills-Higgs systems the
field equations are implied by first-order equations:
\bea
d{=}2{+}2: && \cb D^\mu F_{\mu\nu}=0 \quad\Longleftarrow\quad
F_{\mu\nu}=\sfrac12\eps_{\mu\nu\rho\lambda}F^{\rho\lambda} \non\\
\downarrow \ \quad && \qquad
\cb \downarrow \qquad\qquad\qquad\qquad\qquad \downarrow\\ \non
d{=}2{+}1: && \cb \begin{smallmatrix} D^a F_{ab}\=H D_b H\\
                                           D^a D_a H \=0 \end{smallmatrix}
\quad\Longleftarrow\quad F_{ab}=\eps_{abc}D^c H \ .
\eea
One may gauge-fix and ``solve'' the 3d Bogomolny equations via
\bea
&& \cb A_v \= 0 \qquad\qquad\quad\ \text{and}\qquad\cb 
A_x{+}H \= 0 \ ,\\
&& \cb A_u \= \Phi^{-1}\,\pa_u \Phi \qquad\text{and}\qquad\cb 
A_x{-}H \= \Phi^{-1}\,\pa_x \Phi \ ,
\eea
with  \ $\Phi(u,v,x)\in$ SU(2) \ subject to the ``Yang equation''
\be \cb \label{Yang}
\pa_v ( \Phi^\+ \pa_u \Phi )\ -\ \pa_x ( \Phi^\+ \pa_x \Phi ) \=0\ .
\ee
A dimensional reduction to the 2d WZW model is achieved by letting
\be \cb \label{dimred}
\Phi(u,v,x)\qquad\longrightarrow\qquad
\Ecal\,\e^{\im\a\,x\,\se}\;g(u,v)\,\e^{-\im\a\,x\,\se}\,\Ecal^\+ 
\ee
with a constant matrix $\Ecal$ and \ $g(u,v)\in$ SU(2).
The Yang equation~(\ref{Yang}) then becomes 
\be \cb \label{Yang2}
\pa_v ( g^\+ \pa_u g )\ +\ \a^2 (\se g^\+ \se g - g^\+ \se g\,\se) \=0\ .
\ee
Finally, an algebraic reduction of \ $g(u,v)$ \ to \ U(1) yields
the sine-Gordon equation:
\be \cb
g\=\e^{\frac{\im}{2}\sd\,\p}
\qquad\Longrightarrow\qquad
\pa_v \pa_u \p + 4\a^2 \sin\p \= 0\ .
\ee

\section{Noncommutative deformation}
The Moyal deformation of $\R^{1,1}$ replaces the ordinary pointwise product 
of functions, \\ $(f\!\cdot g)(t,y)=f(t,y)\,g(t,y)$, by the ``star product''
\bea
\cb (f\sta g)(t,y)&\cb\!=\!& \cb f(t,y)\,\exp\,\bigl\{ \sfrac{\im\th}{2}
({\ld{\partial}}_t {\rd{\partial}}_y -
 {\ld{\partial}}_y {\rd{\partial}}_t ) \bigr\}\,g(t,y) \\[4pt]
&\!=\!&f\,g\ +\ \sfrac{\im\th}{2}(\pa_t f\,\pa_y g-\pa_y f\,\pa_t g)
 \ +\ \ldots 
\eea
with a constant noncommutativity parameter $\th\in\R_+$.
This product satisfies, in particular,
\be
(f\,\sta\,g)\,\sta\,h \= f\,\sta\,(g\,\sta\,h)
\qquad\text{and}\qquad
\smallint \diff t\,\diff y\ f\sta g \= \smallint \diff t\,\diff y\, f\,g \ ,
\ee
and the coordinate functions obey the commutation relations
\be
t\,\sta\, y-y\,\sta\, t\ =\ \im\th
\qquad\Longrightarrow\qquad
u\,\sta\, v-v\,\sta\, u\ =\ -\sfrac{\im}{2}\th \ .
\ee
Additional coordinates (for $d{=}2{+}1$ or $d{=}2{+}2$) commute.

\section{Poor deformations of the sine-Gordon model}
Naive $\sta$-ing of the sine-Gordon equation~(\ref{sG}) yields
\be
\cb \pa_v\,\pa_u\,\p \= -4\a^2\,\sin_\sta\p \ ,
\ee
which does not allow for conserved charges.
More promising is the Moyal deformation of the above reduction procedure
(\ref{dimred}), now with \ $\Phi(u,v,x)\in$ U(2) \ and
\be
g(u,v)\=\exp_\sta \bigl\{\sfrac{\im}{2}\sd\,\p(u,v)\bigr\} 
\quad\in\text{U(1)} \ .
\ee
Inserting this into the deformed version of the Yang equation~(\ref{Yang2})
produces {\em two\/} equations,
\bea \cb \label{sG2a}
\pa_v\bigl(\!\es^{-\frac{\im}{2}\p}\,\sta\,\pa_u\es^{\frac{\im}{2}\p} -
             \es^{\frac{\im}{2}\p}\,\sta\,\pa_u\es^{-\frac{\im}{2}\p} \bigr)
&\cb=& \cb -4\im\a^2\sin_\sta\p \ ,\\[2pt] \cb \label{sG2b}
\pa_v\bigl(\!\es^{-\frac{\im}{2}\p}\,\sta\,\pa_u\es^{\frac{\im}{2}\p} +
             \es^{\frac{\im}{2}\p}\,\sta\,\pa_u\es^{-\frac{\im}{2}\p} \bigr)
&\cb=&\cb 0 \ , 
\eea
of which the first one becomes the standard sine-Gordon equation when 
$\th{\to}0$ while the second one may be interpreted as a constraint that 
disappears in the commutative limit.
These equations indeed feature infinitely many conserved local charges,
but the corresponding S-matrix is acausal (containing $\sin^2(pE\th)$ terms)
and yields particle production ($2{\to}3$ \ and \ $2{\to}4$).
Hence, this model does not yet represent a satisfactory deformation of the
sine-Gordon theory.

\section{A proposal: algebraic reduction to \ U(1)$\times$U(1)}
The extension of SU(2) to U(2) for the Yang-Mills gauge group was enforced
by the noncommutativity. It is therefore natural to keep the additional U(1)
factor also in the algebraic reduction. Hence, let me relax the reduction
\be \label{relax} {}\!\!\!\!
\text{from}\qquad g\=\es^{\frac{\im}{2}\sd\,\p} \qquad\text{to}\qquad
g\=\es^{\frac{\im}{2}{\mathbbm{1}}\,\r} \sta \es^{\frac{\im}{2}\sd\,\vp}\ ,
\ee
i.e.~take \  $g(u,v)\in\ $U(1)$\times$U(1) \ and work with {\em two\/}
scalar fields $\vp(u,v)$ and~$\r(u,v)$. 
The Yang equation~(\ref{Yang2}) in this case yields
\bea \label{sG3a}
\pa_v\bigl(\es^{-\frac{\im}{2}\vp}\sta\pa_u\es^{\frac{\im}{2}\vp} \bigr) +
2\im\a^2\sin_\sta\vp &=&
-\pa_v\bigl[\es^{-\frac{\im}{2}\vp}\sta R\sta \es^{\frac{\im}{2}\vp} \bigr] 
\\[2pt] \label{sG3b}
\pa_v\bigl(\es^{\frac{\im}{2}\vp}\sta\pa_u\es^{-\frac{\im}{2}\vp} \bigr) -
2\im\a^2\sin_\sta\vp &=&
-\pa_v\bigl[\es^{\frac{\im}{2}\vp}\sta R\sta \es^{-\frac{\im}{2}\vp} \bigr]
\eea
\vskip-12pt
\be
\text{with}\qquad 
R\ :=\ \es^{-\frac{\im}{2}\r}\sta\pa_u\es^{\frac{\im}{2}\r}\ .
\ee
Note that for $\r=0$ one finds that $R=0$ and recovers 
(\ref{sG2a}) and~(\ref{sG2b}). In the commutative limit $\th{\to}0$ the system
(\ref{sG3a}) plus~(\ref{sG3b}) behaves as it should and decouples to
\be \cb
\pa_v\,\pa_u\,\r \=0 \qquad\text{and}\qquad\cb
\pa_v\,\pa_u\,\vp\ +\ 4\a^2\sin\vp \=0\ .
\ee

\section{Linear system}
In order to unclutter my notation, I suppress all $\sta$ products
for the remainder of the talk but assume their implicit presence
if not said otherwise. 
Therefore, despite appearance even scalar fields do not commute.
Like in the commutative case, also the deformed version of the Yang equation
(\ref{Yang2}) can be seen as the compatibility condition for a 
(now noncommutative) linear system
\bea 
(\pa_u +\im\a\,\z\,\mathrm{ad}\sd)\,\j &=& \ \ -(g^\+\pa_u g)\,\j\ ,\\[2pt]
(\z\,\pa_v +\im\a\,\mathrm{ad}\sd)\,\j &=& \im\a(g^\+\sd\,g)\,\j
\eea
with \ $\j(u,v,\z)\in\text{U(2)}$ \ and limits
\be 
\j(\z{\to}0) \= g^\+ \ +\ O(\z) \qquad\text{and}\qquad 
\j(\z{\to}\infty) \= \unity \ +\ O(\z^{-1}) \ .
\ee
Please note that, due to the Moyal deformation, the entries of all 
these matrices are non-commuting themselves.
In a moment, I am going to exploit the holomorphic dependence on the
spectral parameter $\z\in\C P^1$ in the following three equations:
\bea \label{real}
\unity &=& \j(u,v,\z)\,[\j(u,v,\bar{\z})]^\+ \ ,
\\[4pt] \label{diff1}
g^\+\pa_u g &=& \j\,(\pa_u +\im\a\,\z\,\mathrm{ad}\sd)\,\j^\+ \ ,
\\[4pt] \label{diff2}
-\im\a\,g^\+\sd\,g &=& \j\,(\z\,\pa_v +\im\a\,\mathrm{ad}\sd)\,\j^\+ \ .
\eea
Since $\C P^1$ is compact, a nontrivial (i.e.~non-constant) matrix function
$\j(\z)$ has to be meromorphic. However, the left hand sides of the above 
equations are independent of~$\z$, and so must be their right hand sides.
This fact implies, in particular, that the residues of all poles in the
right-hand-side expressions of (\ref{real}), (\ref{diff1}) and~(\ref{diff2})
better vanish, imposing strong conditions on the auxiliary matrix function 
$\j(u,v,\z)$.

\section{Single-pole ansatz}
The simplest ansatz beyond a constant matrix reads\footnote{
The reason for the seemingly redundant notation becomes clear 
in the next section.}
\be \cb \label{ansatz1}
\j_1 \= \bigl(\unity\,+\,\sfrac{2\im\m_1}{\z-\im\m_1}\,P_1\bigr)\,\j^0_1
\= \bigl(\unity \,+\,\sfrac{\Lambda_{11}S_1^\+}{\z-\im\m_1}\bigr)\,\j^0_1
\ee
with $\m_1\in\R$ (an imaginary pole) 
and a constant matrix $\j^0_1\in\text{U(2)}$.
To be determined are the U(2) valued noncommutative functions 
\ $P_1(u,v)$ \ and \ $\Lambda_{11}S_1^\+(u,v)$.
Inserting the ansatz (\ref{ansatz1}) into (\ref{real})
and isolating the residues one gets 
\be \label{res1R}
\res_{\z=-\im\m_1}(\ref{real})=0 \quad\Longrightarrow\quad
\begin{cases}
{}\quad P_1^\+ \= P_1 \= P_1^2 \quad&\Longrightarrow\quad\cb
P_1 \= T_1\,\sfrac{1}{T_1^\+ T_1}\,T_1^\+ \\[4pt] \cb
(\unity{-}P_1)\,S_1\Lambda_{11}^\+ \=0 \quad&\Longrightarrow\quad\cb
T_1 \= S_1
\end{cases}
\ee
which qualifies $P_1$ as a hermitian projector built from a $2{\times}1$ 
matrix~$T$ which spans~im$P_1$. 
Next, exploiting (\ref{diff1}) and~(\ref{diff2}) yields
\be \label{res1D}
\res_{\z=-\im\m_1}(\ref{diff1},\ref{diff2})=0 \quad\Longrightarrow\quad
(\unity{-}P_1)\,\Lb_1\,(S_1\Lambda_{11}^\+) \=0
\quad\Longrightarrow\cb\quad
\Lb_1\,S_1 \= S_1\,\Ga_1
\ee
with a constant $\Ga_1$ and
\be
\Lb_i \ :=\ \begin{cases}
\qquad \pa_u +\a\,\mu_i\,\mathrm{ad}\sd 
\qquad &\text{for}\quad (\ref{diff1}) \\[4pt]
-\mu_i^2\,\pa_v +\a\,\mu_i\,\mathrm{ad}\sd 
\qquad &\text{for}\quad (\ref{diff2}) \end{cases} 
\qquad\qquad(\text{here}\quad i=1) \ .
\ee
The residues at $\z=\im\m_1$ merely lead to the complex conjugated conditions.
The solution to (\ref{res1D}) has the form 
(I choose \ $\g_{11}, \g_{12} \in\R$)
\be \cb \label{solS1}
S_1(u,v) \= \widehat{S}_1(\h_1) \= \e^{-\a\,\h_1\sd}\,
\bigl(\begin{smallmatrix} \g_{11} \\[2pt] \im\g_{12} \end{smallmatrix}\bigr)
\ee
and combines the $u$, $v$ dependence in a single ``co-moving coordinate''
\be \label{comov}
\h_i\=\m_i u -\sfrac{1}{\m_i} v \=\sfrac{y-\vel_i t}{\sqrt{1-\vel_i^2}} 
\qquad\qquad (\text{here}\quad i=1)\ .
\ee

\section{Dressing method}
I proceed to the two-pole ansatz, in a multiplicative and an additive form:
\bea \cb \label{ansatz2a}
\j_2 &\cb=&\cb \bigl(\unity\,+\,\sfrac{2\im\m_2}{\z-\im\m_2}\,P_2\bigr)
\bigl(\unity\,+\,\sfrac{2\im\m_1}{\z-\im\m_1}\,P_1\bigr) \,\j^0_2 
\\[4pt] \cb \label{ansatz2b}
&\cb=&\cb \bigl(\unity\,+\,\sfrac{\Lambda_{21}S_1^\+}{\z-\im\m_1}
\,+\,\sfrac{\Lambda_{22}S_2^\+}{\z-\im\m_2}\bigr) \,\j^0_2 \ ,
\eea
generalizing the one-pole notation of (\ref{ansatz1}) in an obvious way.
A look at the residues at $\z=-\im\m_1$ reveals that $P_1$ and $S_1$ 
are subject to the same equations (\ref{res1R}) and (\ref{res1D}) as in the 
one-pole case and thus can be taken over from there, e.g.~via~(\ref{solS1}).
The analysis of the residues at $\z=-\im\m_2$ is more involved however.
First, the two forms (\ref{ansatz2a}) and (\ref{ansatz2b}) yield
\vskip-24pt
\be \label{res2R}
\res_{\z=-\im\m_2}(\ref{real})=0 \quad\Longrightarrow
\Biggl\{ \begin{matrix} {}\\[4pt]
(\unity{-}P_2)\,P_2\=0 \quad\Longrightarrow\quad\cb
P_2 \= T_2\,\sfrac{1}{T_2^\+ T_2}\,T_2^\+ \qquad \\[8pt]
\cb \j_2(\m_2)\,S_2\Lambda_{22}^\+ \=
(\unity{-}P_2)\underbrace{(1-\sfrac{2\m_1}{\m_1+\m_2}P_1)\,S_2}_{T_2}
\Lambda_{22}^\+\=0 
\end{matrix}
\ee
\vskip-12pt \noindent
and, second, the additive variant (\ref{ansatz2b}) produces
\be \label{res2D}
\res_{\z=-\im\m_2}(\ref{diff1},\ref{diff2})=0 \quad\Longrightarrow\quad
\j_2(\m_2)\,\Lb_2\,(S_2\Lambda_{22}^\+) \=0
\quad\Longrightarrow\cb\quad
\Lb_2\,S_2 \= S_2\,\Ga_2 
\ee
with a constant $\Ga_2$.
Like before, the solution to the latter equation reads
(take \ $\g_{21}, \g_{22} \in\R$)
\be \cb
S_2(u,v) \= \widehat{S}_2(\h_2) \= \e^{-\a\,\h_2\sd}\,
\bigl(\begin{smallmatrix} \g_{21} \\[2pt] \im\g_{22} \end{smallmatrix}\bigr)
\ee
with a second co-moving coordinate $\h_2$ already defined in (\ref{comov}).

The iteration of this dressing procedure to the construction of
higher-pole solutions $\j_N$ is now straightforward. The strategy is to
choose pole locations $\m_i$ (or velocities $\vel_i$) and real constants 
$\g_{ik}$ and then rebuilt recursively in the order
\be
\m_i,\g_{ik}\ \to\ S_i\ \to\ T_i\ \to\ P_i\ \to\ \j_N\ \to\ g_N
\qquad\text{for}\quad i=1,\dots,N \ .
\ee

\section{Noncommutative kinks}
The $N$-pole solutions produced with the dressing method just outlined
turn out to be noncommutative multi-solitons, i.e.~they possess finite energy
and approach their commutative cousins for $\th{\to}0$. Let me elaborate
on the simplest case, $N=1$:
\be \cb
S(u,v) \= \e^{-\a\,\h\,\sd}\,
\bigl(\begin{smallmatrix} \g_{1} \\[2pt] \im\g_{2} \end{smallmatrix}\bigr)
\= \sqrt{|\g_1\g_2|}\; \e^{-\a\,(\h-\h_0) \sd}\,
\bigl(\begin{smallmatrix} 1 \\[2pt] \im \end{smallmatrix}\bigr)\ ,
\ee
where $\h_0=\frac{1}{2\a}\ln|\frac{\g_1}{\g_2}|$ determines the center of mass
at $t{=}0$. For simplicity I put $\h_0=0$ and calculate
\be \cb \label{onesol}
T = \biggl(\begin{matrix} \!\e^{-\a\h} \\[2pt] 
                          \!\im\e^{\a\h} \end{matrix}\biggr) 
\quad\Rightarrow\quad
P = \frac{1}{2\,\ch 2\a\h} \biggl(\begin{matrix}
\!\e^{-2\a\h} \!\! & -\im \\[4pt] \im & \!\! \e^{+2\a\h} \end{matrix}\biggr) 
\quad\Rightarrow\quad
g = \biggl(\begin{matrix} \tah 2\a\h & \frac{\im}{\ch 2\a\h}\\[4pt]
                \frac{\im}{\ch 2\a\h} & \tah 2\a\h \end{matrix}\biggr) \ .
\ee
Since the $u$, $v$ dependence resides only in the single coordinate~$\h$,
all $\sta$ products trivialize and one effectively falls back on the
$\th{=}0$ (and hence $\r{=}0$) situation.
Comparing $g$ of (\ref{onesol}) with the form introduced in (\ref{relax}),
for $\r{=}0$ and modulo an admissible constant rotation,
\be
g \= \e^{\frac{\im}{2}\se\,\vp} \= \Ecal\,\e^{\frac{\im}{2}\sd\,\vp}\,\Ecal^\+
\qquad\text{for}\quad \Ecal = \e^{-\im\frac{\pi}{4}\sz} \ ,
\ee
one reads off that
\be \cb
\cos\sfrac{\vp}{2} \= \tah 2\a\h
\qquad\textrm{and}\qquad
\sin\sfrac{\vp}{2} \= \sfrac{1}{\ch 2\a\h}
\qquad\Longrightarrow\qquad
\tan\sfrac{\vp}{4} \= \e^{-2\a\h} \ ,
\ee
which indeed yields the standard sine-Gordon kink~(\ref{kink}) with
velocity \ $v=\sfrac{1-\m^2}{1+\m^2}$ \ but no deformation.
Note, however, that breathers and multi-solitons will get deformed
because different co-moving coordinates do not commute,
\be
[\h_i\,,\h_k]\=
-\im\th\;\frac{\vel_i-\vel_k}{\sqrt{(1{-}\vel_i^2)(1{-}\vel_k^2)}}\ .
\ee

\section{Tree-level S-matrix}
The noncommutative sine-Gordon equations (\ref{sG3a}) and (\ref{sG3b})
also follow from an action principle, which allows for a quick derivation
of the Feynman rules. The two scalars $\vp$ and $\r$ have masses $2\a$ and 0,
respectively, and are coupled via an infinite sequence of 
higher-derivative interactions. As a constructive example I consider the
$\vp\vp\to\vp\vp$ tree-level scattering amplitude, with the kinematics
($E^2-p^2=4\a^2$)
\be
k_1=(E,p)\ ,\quad k_2=(E,-p)\ ,\quad k_3=(-E,p)\ ,\quad k_4=(-E,-p) \ .
\ee
The sum of the relevant four-point diagrams
\begin{flushleft}
\begin{tabular}{ccccccc}  
\parbox{1.9cm}{\includegraphics[width=1.9cm]{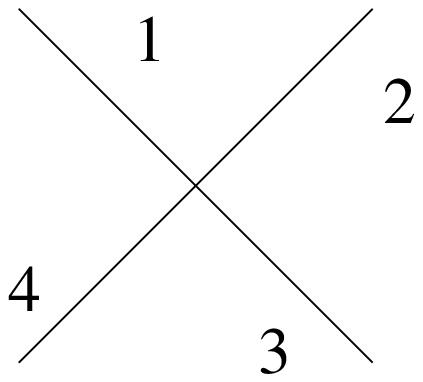}} & $\!{+}\!$ &
\parbox{1.9cm}{\includegraphics[width=1.9cm]{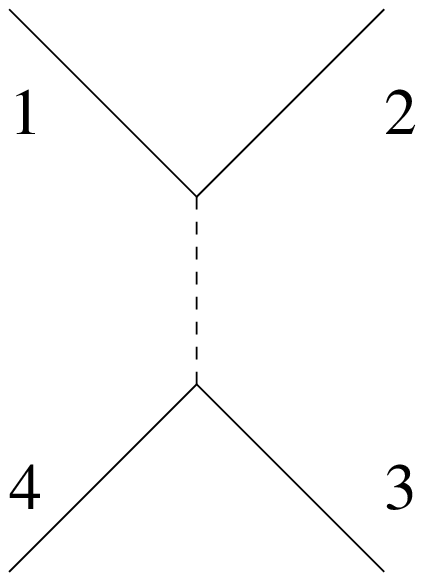}} & $\!{+}\!$ &
\parbox{2.4cm}{\includegraphics[width=2.4cm]{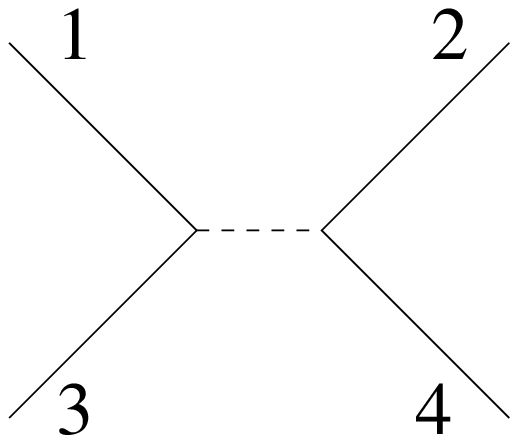}} & $\!{+}\!$ &
\parbox{2.4cm}{\includegraphics[width=2.4cm]{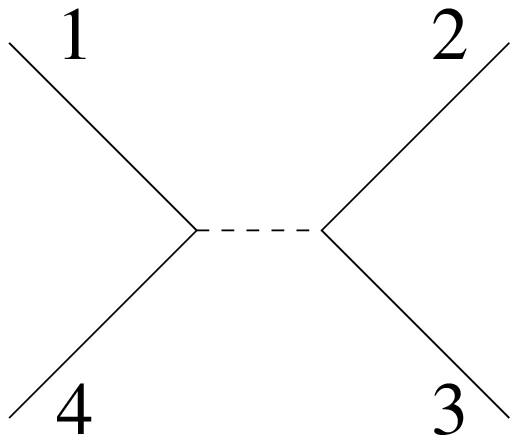}} \\ 
&&&&&& \\
$2\!\im\!\a^2\!\cos^2 (\theta Ep)$     & $\!{-}\!$ &
$\frac{\im}{2}p^2\!\sin^2 (\theta Ep)$ & $\!{+}\!$ &
$\frac{\im}{2}E^2\!\sin^2 (\theta Ep)$ & $\!{+}\!$ & $0$
\end{tabular}
\end{flushleft}
add up to \ ${\cb A_{\vp\vp\to\vp\vp}\=2\im\a^2}$ \ which is causal!
Likewise, one can show that all other $2{\to}2$ amplitudes vanish. Also,
\ ${\cb\vp\vp\to\vp\vp\vp\vp}$ \ and \ ${\cb\vp\vp\vp\to\vp\vp\vp}$ \ 
do not occur, indicating the absence of particle production. 
The S-matrix appears to be causal and factorizable at tree level.

Most results presented here and all relevant references can be found
in~\cite{paper}.

\end {document}